\documentclass[aoas,nameyear,dvips]{arximspdf}
\usepackage{accents}
\usepackage{graphicx}

\doi{10.1214/10-AOAS355}
\volume{4}
\issue{4}
\pubyear{2010}
\firstpage{2150}
\lastpage{2180}

\makeatletter
\newtheorem{proposition}{Proposition}
\newproclaim{Rem}{Remarks}
\makeatother

\begin{document}
\begin{frontmatter}

\title{Sparse modeling of categorial explanatory variables}
\runtitle{Sparse modeling of categorial explanatory variables}

\begin{aug}
\author[A]{\fnms{Jan} \snm{Gertheiss}\thanksref{t1}\ead[label=e1]{jan.gertheiss@stat.uni-muenchen.de}\corref{}}
\and
\author[A]{\fnms{Gerhard} \snm{Tutz}\ead[label=e2]{gerhard.tutz@stat.uni-muenchen.de}}
\thankstext{t1}{Supported in part by DFG project TU62/4-1 (AOBJ: 548166).}
\runauthor{J. Gertheiss and G. Tutz}
\affiliation{Ludwig-Maximilians-Universit\"{a}t Munich}
\address[A]{Department of Statistics\\
Ludwig-Maximilians-Universit\"{a}t Munich\\
Akademiestrasse 1\\
80799 Munich\\
Germany\\
\printead{e1}\\
\phantom{E-mail:\ }\printead*{e2}} 
\end{aug}

\received{\smonth{6} \syear{2009}}
\revised{\smonth{4} \syear{2010}}

\begin{abstract}
Shrinking methods in regression analysis are usually designed for
metric predictors. In this article, however, shrinkage methods for
categorial predictors are proposed. As an application we consider
data from the Munich rent standard, where, for example, urban
districts are treated as a categorial predictor. If independent
variables are categorial, some modifications to usual shrinking
procedures are necessary. Two $L_1$-penalty based methods for factor
selection and clustering of categories are presented and
investigated. The first approach is designed for nominal scale
levels, the second one for ordinal predictors. Besides applying them
to the Munich rent standard, methods are illustrated and compared in
simulation studies.
\end{abstract}


\end{frontmatter}

\section{Introduction}
Within the last decade regularization, and in particular variable
selection, has been a topic of intensive research. With the
introduction of the Lasso, proposed by \citet{Tibshirani96}, sparse
modeling in the high-dimensional predictor case with good
performance, in terms of identification of relevant variables
combined with good performance in predictive power, became possible.
In the following many alternative regularized estimators that
include variable selection were proposed, among them the Elastic Net
[\citet{ZouHas2005}], SCAD [\citet{FanLi2001}], the Dantzig Selector
[\citet{CanTao2007}] and Boosting approaches [for example,
\citet{BueYu2003}].

This article provides a regularized regression analysis of Munich
rent standard data. All larger German cities publish so-called rent
standards for having guidelines available to tenants, landlords,
renting advisory boards and experts. These rent standards are used,
in particular, to determine the local comparative rent. For the
composition of rent standards, a representative random sample is
drawn from all relevant households, and the interesting data are
determined by interviewers by means of questionnaires. The data
analyzed come from 2053 households interviewed for the Munich rent
standard 2003. The response is monthly rent per square meter in
Euro. The predictors are ordered as well as unordered and binary
factors. A detailed description is given in Table
\ref{TableRental1}. The data can be downloaded from the data archive
of the Department of Statistics at the University of Munich
(\url{http://www.stat.uni-muenchen.de/service/datenarchiv}). The direct
link is found
there.

For example, the urban district is given as a nominal predictor with
25 possible values. The decade of construction can be interpreted as
ordinal with 10 levels. Usually such data are analyzed via standard
linear regression modeling, with (for example) dummy coded
categorial explanatory variables. In the present situation such
modeling is possible, since the number of observations (2053) is
quite high. Nevertheless, from the viewpoint of interpretation, model
selection is desired with the focus on reducing model complexity.

In selection problems for categorical predictors as in the Munich
rend data example, it should be distinguished between two problems:
\begin{itemize}
\item Which categorical predictors should be included in the model?
\item Which categories within one categorical predictor should be
distinguished?
\end{itemize}
The latter problem is concerned with one variable and poses the
question of which categories differ from one another with respect to
the dependent variable. Or, to put it in a different way, which
categories should be collapsed? The answer to that question depends
on the scale level of the predictor, one should distinguish between
nominal and ordered categories because of their differing
information content.

\begin{table}
\caption{\label{TableRental1}Explanatory variables for monthly rent
per square meter}
\begin{tabular*}{\textwidth}{@{\extracolsep{\fill}}ll@{}}
\hline
Urban district & Nominal, labeled by numbers\\
 & $1,\ldots,25$\\[5pt]
Year of construction & Given in ordered classes $[1910,1919],$\\
 & $[1920,1929], \ldots$\\
Number of rooms & Taken as ordinal factor with levels\\
 & $1,2,\ldots,6$\\
Quality of residential area & Ordinal, with levels ``fair,''\\
 & ``good,'' ``excellent''\\
Floor space (in $\mbox{m}^2$) & Given in ordered classes $(0,30),$\\
 & $[30,40)$, $[40,50)$, $\ldots$ , $[140,\infty)$\\[5pt]
Hot water supply & Binary (yes/no)\\
Central heating & Binary (yes/no)\\
Tiled bathroom & Binary (yes/no)\\
Supplementary equipment in bathroom & Binary (no/yes)\\
Well equipped kitchen & Binary (no/yes)\\
\hline
\end{tabular*}
\end{table}

When investigating which of the 25 urban districts of Munich are to
be distinguished with respect to the local rent, the number of
possible combinations is huge. If only urban districts are used as
categorial predictor in a regression model to explain the monthly
rent, and districts are potentially fused (without further
restrictions), the number of possible models---which just follow
from different fusion results---is greater than $10^{18}$. In cases
like that---that is, when the number of possible models is large---regularization techniques which induce sparsity are a promising
approach for model selection. The extent of regularization---and
hence sparsity---is typically controlled by a tuning parameter. Via
choosing this parameter, the model is also implicitly selected.

Most of the regularization techniques developed so far focus on the
selection of variables in the case where the effect of one variable
is determined by one coefficient. That means coefficients are
selected rather than variables. When all predictors are metric and a
main effect model is assumed to hold, of course selection of
coefficients is equivalent to selection of predictor variables and
model selection. This is different when categorical variables have
to be included because then a whole group of coefficients refers to
one variable.

To be more concrete, let us first consider just one categorial
predictor $C \in \{0,\ldots,k\}$ and dummy coding $x_i =
I_{\{C=i\}}$. Then the classical linear model is given as
\[
y = \alpha + \sum_{i=0}^k \beta_ix_i + \epsilon,
\]
with $E(\epsilon) = 0$ and $\mbox{Var}(\epsilon) = \sigma^2$. If
category $0$ is chosen as reference, coefficient $\beta_0$ is fixed
to zero. When computing a penalized estimate, for example, by use of
the simple Lasso [\citet{Tibshirani96}], the shrinkage effect depends
on the coding scheme that is used and the choice of the reference
category. With category zero chosen as reference, shrinkage always
refers to the difference between category $i$ and zero. Moreover,
Lasso type penalties tend to set some coefficients to zero. Usually
this feature is seen as a great advantage over methods like Ridge
regression, since it can be used for model/variable selection.
Applied to dummy coded categorial predictors, however, selection
only refers to the currently chosen reference category. In most
cases of nominal predictors, class labeling and choice of the
reference category is arbitrary, which means that the described
selection procedures are not really meaningful. In addition, the
estimated model is not invariant against irrelevant permutations of
class labels.

\begin{figure}

\includegraphics{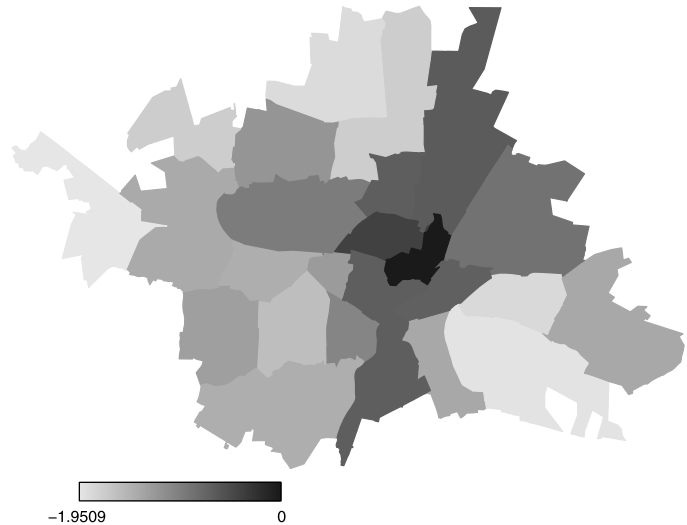}

\caption{Map of Munich indicating urban
districts; colors correspond to estimated dummy coefficients if an
ordinary least squares model is fitted with predictors from Table
\protect\ref{TableRental1} and response monthly rent per square meter (in
Euro).}\label{FigureRental1a}
\end{figure}

One of the few approaches that explicitly select categorical
predictors was proposed by \citet{YuanLin2006} under the name
\textsl{Group Lasso}. The approach explicitly includes or excludes
groups of coefficients that refer to one variable. However, while
the Group Lasso only attacks the problem of factor selection, for
categorical predictor variables with many categories a useful
strategy is to (additionally) search for clusters of categories with
similar effects. As already described, in the presented application
(among other things) we try to model the influence of the urban
district where a person lives on the rent she/he has to pay. In
Figure \ref{FigureRental1a} a map of Munich is drawn with color
coded urban districts. Colors correspond to dummy coefficients if an
ordinary least squares model is fitted with (dummy coded)
explanatory variables from Table \ref{TableRental1} and response
monthly rent per square meter (in Euro). Some districts are hard to
distinguish. That means it can be expected that not all districts do
differ substantially. If an ordinary least squares model is fitted,
however, estimated dummy coefficients (almost surely) differ.
Therefore, the aim is to combine districts which (on average) do not
substantially differ in terms of rent per square meter. Generally
speaking, that means the objective is to reduce the $k+1$ categories
to a smaller number of categories which form clusters. The effect of
categories within one cluster is supposed to be the same but
responses will differ across clusters. Therefore, in a regression
model corresponding dummy coefficients should be equal. Since,
however, the number of possible clustering results---and hence the
number of models---tends to be very large (as already mentioned),
model selection via regularization is quite attractive.

\begin{figure}

\includegraphics{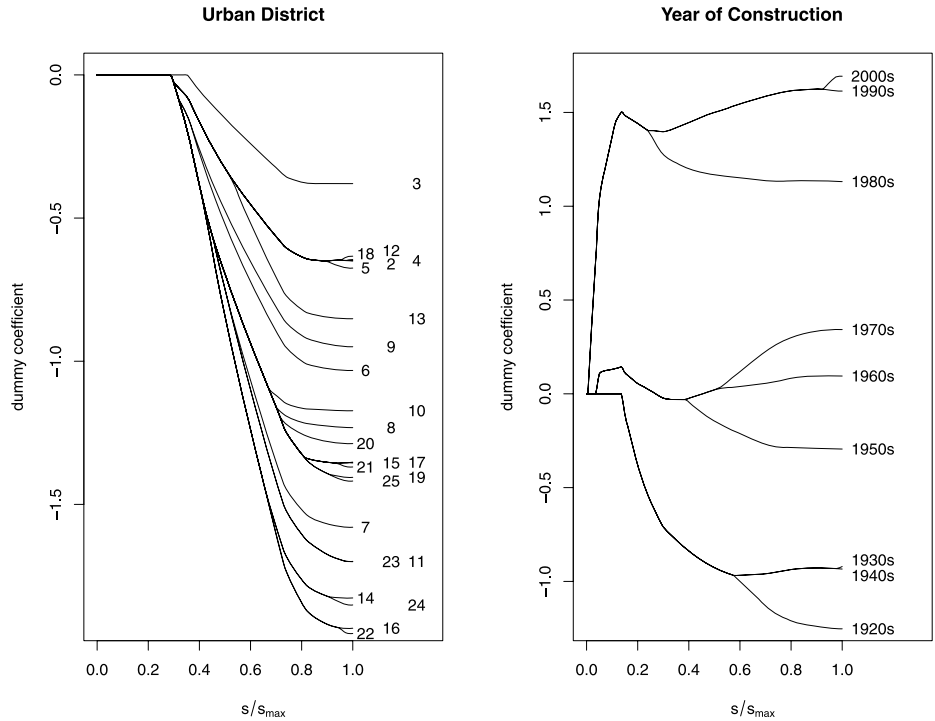}

\caption{Paths of dummy coefficients of two
categorial predictors obtained by the regularization technique
proposed here for the analysis of rent standard data.}\label{FigureRental1b}
\end{figure}

Clustering or \textsl{fusion} of metric predictors may, for example,
be obtained by so-called Variable Fusion [\citet{LanFri1997}] and the
Fused Lasso proposed by \citet{TibSauRosZhuKni2005}. If predictors
can be ordered, by putting a $L_1$-penalty on differences of
adjacent coefficients many of these differences are set to zero,
yielding a piecewise constant coefficient function. Recently,
\citet{BonRei2009} adapted this methodology for factor selection
and level fusion in ANOVA, to obtain dummy coefficients that are
constant over some of the categories. The main focus of
\citet{BonRei2009}, however, was on ANOVA typical identification of
differences, not on model building as in our case, where
prediction accuracy is also an important aspect. So in the following the
method is reviewed and adapted to regression type problems. Some
modifications are proposed and an approximate solution is presented
which allows for easy computation of coefficient paths. In addition,
the method is adapted to the modeling of ordinal predictors.

Figure \ref{FigureRental1b} shows paths of dummy coefficients for
the rent data obtained by the method used in this article. The
coefficients at value $s/s_{\max} = 1$ correspond to the ordinary
least squares model. It is seen that with decreasing tuning
parameter $s$, categories are successively fused, that is,~coefficients
are set equal. Besides the urban district, several other covariates
are given, among them the (categorized) year of construction.
Corresponding paths of dummy coefficients are also shown in Figure
\ref{FigureRental1b}.

\section{Regularization for categorical predictors}
In the following we consider the penalized least squares criterion
\begin{equation}\label{DefQp}
Q_p(\beta) = (y-X\beta)^T(y-X\beta) + \lambda J(\beta),
\end{equation}
with design matrix $X$, coefficient vector $\beta$ and penalty
$J(\beta)$; $y$ contains the observed response values. The estimate
of $\beta$ is given by
\begin{equation}\label{Defbetahat}
\hat\beta = \mathop{\operatorname{argmin}}_\beta \{Q_p(\beta)\}.
\end{equation}
The decisive point is a suitable choice of penalty $J(\beta)$. We
start with the case of one categorial explanatory variable and will
distinguish between nominal and ordinal predictors.

\subsection{Unordered categories}
If the categorial predictor has only nominal scale level,  a
modification of Variable Fusion [\citet{LanFri1997}] and the Fused
Lasso [\citet{TibSauRosZhuKni2005}] has been proposed by
\citet{BonRei2009} in the form of the penalty
\begin{equation}\label{DefPenCat}
J(\beta) = \sum_{i>j} w_{ij}|\beta_i - \beta_j|,
\end{equation}
with weights $w_{ij}$ and $\beta_i$ denoting the coefficient of
dummy $x_i$. Since the ordering of $x_0,\ldots,x_k$ is arbitrary,
not only differences $\beta_i - \beta_{i-1}$ (as in original fusion
methodology), but all differences $\beta_i - \beta_j$ are
considered. Since $i=0$ is chosen as reference, $\beta_0 = 0$ is
fixed. Therefore, in the limit case, $\lambda \rightarrow \infty$,
all $\beta_i$ are set to zero and the categorial predictor $C$ is
excluded from the model since no categories are distinguished
anymore. For $\lambda < \infty$ the Lasso type penalty
(\ref{DefPenCat}) sets only some differences $\beta_i - \beta_j$ to
zero, which means that categories are clustered. With adequately
chosen weights $w_{ij}$, some nice asymptotic properties like
selection and clustering consistency of $\hat\beta$ can be derived.
These (adaptive) weights decisively depend on the distance of the
ordinary least squares estimates $\hat\beta_i^{(LS)}$ and
$\hat\beta_j^{(LS)}$. For details see Proposition \ref{PropCat} in
the \hyperref[app]{Appendix}. The issue, how to select concrete weights in
the $n < \infty$ case, is further addressed in Sections
\ref{SectionWeights} and \ref{SectionSim2}.

\subsection{Ordered categories}
An interesting case are selection strategies for ordinal predictors,
as, for example, the decade of construction from Table
\ref{TableRental1}. Ordered categories contain more information than
unordered ones, but the information has not been used in the
penalties considered so far. Since in the case of ordered categories
the ordering of dummy coefficients is meaningful, original fusion
methodology can be applied, which suggests penalty
\begin{equation}\label{DefPenOrd}
J(\beta) = \sum_{i=1}^k w_i |\beta_i - \beta_{i-1}|,
\end{equation}
with $\beta_0 = 0$. In analogy to asymptotic properties for the
unordered case, with adequately chosen weights $w_i$, similar
results can be derived; see the \hyperref[app]{Appendix} for details.

\subsection{Computational issues}\label{SectionComp}
For the actual application of the proposed method a fitting
algorithm is needed. For that purpose it is useful to consider the
penalized minimization problem (\ref{Defbetahat}) as a constrained
minimization problem. That means $(y-X\beta)^T(y-X\beta)$ is
minimized subject to a constraint. For unordered categories the
constraint corresponding to penalty (\ref{DefPenCat}) is
\[\sum_{i>j} w_{ij} |\beta_i - \beta_j| \le s,\]
with $\beta_0 = 0$. There is a one-to-one correspondence between the
bound $s$ and penalty parameter $\lambda$ in (\ref{DefQp});
cf.~\citet{BonRei2009}. For estimation purposes we consider
transformed parameters $\theta_{ij} = \beta_i - \beta_j$ which yield
vector $\theta = (\theta_{10},\theta_{20},\ldots,\theta_{k,k-1})^T.$
If $\theta$ is directly estimated (instead of $\beta$), one has to
take into account that restrictions $\theta_{ij} = \theta_{i0} -
\theta_{j0}$ must hold for all $i,j > 0$. For practical estimation,
parameters $\theta_{ij}$ are additionally split into positive and
negative parts, that is,
\[\theta_{ij} = \theta_{ij}^+ - \theta_{ij}^-,\]
with
\[\theta_{ij}^+ \ge 0,\qquad \theta_{ij}^- \ge 0,\]
and
\[\sum_{i>j} w_{ij}(\theta_{ij}^+ + \theta_{ij}^-) \le s.\]
Minimization can be done by using quadratic programming methods. We
used R 2.9.0 [\citet{RDev2009}] and the interior point optimizer from
add-on package \texttt{kernlab} [\citet{KarSmoHorZei2004}].

The problem with quadratic programming is that the solution can only
be computed for a single value $s$. To obtain a coefficient path (as
in Figure \ref{FigureRental1b}), the procedure needs to be applied
repeatedly. Moreover, when applying the method to our data, we found
numerical problems, especially when $s$ was small. To attack these
problems, we propose an approximate solution which can be computed
using R add-on package \texttt{lars} [\citet{Efronetal2004}], where
``approximate'' means that only $\theta_{ij} \approx \theta_{i0} -
\theta_{j0}$ holds. For simplicity, we assume that weights $w_{ij} =
1$ are chosen. But results can be generalized easily (see Section
\ref{SectionWeights}). For the approximation we exploit that the
proposed estimator can be seen as the limit of a generalized Elastic
Net. The original Elastic Net [\citet{ZouHas2005}] uses a combination
of simple Ridge and Lasso penalties. We use a generalized form where
the quadratic penalty term is modified. With $Z$ so that $Z\theta =
X\beta$, we define
\[
\hat\theta_{\gamma,\lambda} = \mathop{\operatorname{argmin}}_{\theta}
\biggl\{(y-Z\theta)^T(y-Z\theta) + \gamma\sum_{i>j>0} (\theta_{i0} -
\theta_{j0} - \theta_{ij})^2 + \lambda \sum_{i>j}
|\theta_{ij}|\biggr\}.
\]
A simple choice of $Z$ is $Z =(X|0)$, since
$\theta_{i0} = \beta_i$, $i=1,\ldots,k$. The first penalty term,
which is weighted by $\gamma$, penalizes violations of restrictions
$\theta_{ij} = \theta_{i0} - \theta_{j0}$. The exact solution of the
optimization problem considered here is obtained as the limit
\[\hat\theta = \lim_{\gamma \rightarrow \infty} \hat\theta_{\gamma,\lambda}.\]
Hence, with sufficiently high $\gamma$, an acceptable approximation
should be obtained. If matrix $A$ represents restrictions
$\theta_{ij} = \theta_{i0} - \theta_{j0}$ in terms of $A\theta = 0$,
one may define precision by
\[
\Delta_{\gamma,\lambda} =
(A\hat\theta_{\gamma,\lambda})^TA\hat\theta_{\gamma,\lambda}.
\]
The
lower $\Delta_{\gamma,\lambda}$ the better. An upper bound is given
by
\[
\Delta_{\gamma,\lambda} \le \frac{\lambda (|\hat\theta^{(LS)}| - |\hat\theta_{0,\lambda}|)}{\gamma},
\]
where $\hat\theta^{(LS)}$ denotes the least squares estimate
(i.e., $\lambda = 0$) where $A\hat\theta^{(LS)}=0$ holds, and
$|\theta| = \sum_{i>j} |\theta_{ij}|$ denotes the $L_1$-norm of
vector $\theta$. (For a proof see the \hyperref[app]{Appendix}.)
$\hat\theta^{(LS)}$ can be computed by $\hat\theta_{\gamma,0}$ if
any $\gamma > 0$ is chosen. Not surprisingly, for higher $\lambda$
higher $\gamma$ must also be chosen to stabilize precision.

The advantage of using the estimate $\hat\theta_{\gamma,\lambda}$ is
that its whole path can be computed using \texttt{lars}
[\citet{Efronetal2004}], since it can be formulated as a Lasso
solution. With augmented data $\widetilde Z =
(Z^T,\sqrt{\gamma}A^T)^T$ and $\widetilde y = (y^T, 0)^T$, one has
\[\hat\theta_{\gamma,\lambda} = \mathop{\operatorname{argmin}}_{\theta}
\biggl\{(\widetilde y - \widetilde Z \theta)^T(\widetilde y -
\widetilde Z \theta) + \lambda \sum_{i>j} |\theta_{ij}|\biggr\},\]
which is a Lasso type problem on data $(\widetilde y, \widetilde
Z)$.

In the case of ordinal predictors the penalty is
\[
J(\beta) = \sum_{i=1}^k |\beta_i- \beta_{i-1}|,
\]
and the corresponding optimization problem can be
directly formulated as a simple Lasso type problem. We write
\[
Q_p(\beta) = (y-X\beta)^T(y-X\beta) + \lambda J(\beta) =
(y-\widetilde X\delta)^T(y- \widetilde X\delta) + \lambda
J(\delta),
\]
with $\widetilde X = XU^{-1}$, $\delta = U\beta$,
$J(\delta) = \sum_{i=1}^k |\delta_i|$, and
\[
U = \pmatrix{
1 & 0 & \cdots & 0 \cr
-1 & 1 & \cdots & 0 \cr
0 & \ddots & \ddots & 0 \cr
0 & \cdots & -1 & 1 \cr
}.
\]
Simple matrix multiplication shows that the inverse of $U$ is given
by
\[
U^{-1} = \pmatrix{
1 & 0 & \cdots & 0 \cr
1 & 1 & \ddots & \vdots \cr
\vdots &  &\ddots & 0 \cr
1 & \cdots & \cdots & 1}.
\]
In other words, the ordinal input is just split-coded
[\citet{WalFeiWel87}], and ordinary Lasso estimation is applied.
Split-coding means that dummies $\widetilde x_i$ are defined by
splits at categories $i=1,\ldots,k$, that is,
\[
\widetilde x_i = \cases{ 1, &\quad \mbox{if } $C \ge i$,\cr
0, &\quad \mbox{otherwise}.}
\]
Now the model is parameterized by coefficients $\delta_{i} =
\beta_{i} - \beta_{i-1}$, $i=1,\ldots,k$. Thus, transitions between
category $i$ and $i-1$ are expressed by coefficient $\delta_i$.
Original dummy coefficients are obtained by back-transformation
$\beta_i = \sum_{s=1}^i \delta_s$. By applying penalty $\sum_{i=1}^k
|\delta_i|$, not the whole ordinal predictor is selected, but only
relevant transitions between adjacent categories. By contrast,
\citet{WalFeiWel87} intended the use of classical tests for such
identification of substantial ``\textsl{between-strata differences}.''

\subsection{Multiple inputs}
In our application, as usual in statistical modeling, a set of
(potential) regressors is available (see Table \ref{TableRental1})
and only the relevant predictors should be included into the model.
In the introduction we already considered two predictors, the urban
district where a flat is located and the decade of construction. For
the handling of multiple categorial predictors in general, say,
$x_1,\ldots,x_p$, with levels $0,\ldots,k_l$ for variable $x_l$
($l=1,\ldots,p$, and fixed $p$), the presented methods can be easily
generalized. The corresponding penalty is
\begin{equation}\label{DefMultPen}
 J(\beta) = \sum_{l=1}^p J_l(\beta_l),
\end{equation}
with
\[
J_l(\beta_l) = \sum_{i>j} w_{ij}^{(l)} |\beta_{li} -
\beta_{lj}|, \quad\mbox{or}\quad J_l(\beta_l) = \sum_{i=1}^{k_l} w_{i}^{(l)}
|\beta_{li} - \beta_{l,i-1}|,
\]
depending on the scale level of
predictor $x_l$. The first expression refers to nominal covariates,
the second to ordinal ones.

If multiple predictors are considered, clustering of categories of
single predictors as well as selection of predictors is of interest.
Penalty (\ref{DefMultPen}) serves both objectives, clustering and
selection. If all dummy coefficients that belong to a specific
predictor are set to zero, the corresponding predictor is excluded
from the model. Within each nominal predictor $x_l$, there is also
an $L_1$-penalty on the differences to the dummy coefficient of the
reference category. Since the latter is fixed to zero, clustering of
all categories of $x_l$ means that all coefficients which belong to
predictor $x_l$ are set to zero. In the ordinal case, this happens
if all differences $\delta_{li} = \beta_{li} - \beta_{l,i-1}$ of
adjacent dummy coefficients of predictor $x_l$ are set to zero.

\subsection{Incorporation of weights}\label{SectionWeights}
In many situations weights $w_{ij}^{(l)} \neq 1$ are to be preferred
over the simple weights $w_{ij}^{(l)} = 1$, for example, to obtain
the adaptive versions described in Propositions \ref{PropCat} and
\ref{PropOrd} in the \hyperref[app]{Appendix}, or when predictors differ in
the number of levels, as in the rent standard application (see Table
\ref{TableRental1}). For nominal variables \citet{BonRei2009}
suggested the weights
\begin{equation}\label{DefWeights}
w_{ij}^{(l)} = (k_l + 1)^{-1} \sqrt{\frac{n_i^{(l)} +
n_j^{(l)}}{n}},
\end{equation} where $n_i^{(l)}$ denotes the number of observations
on level $i$ of predictor $x_l$. In the adaptive version the weights
contain additionally the factor $|\hat\beta_{li}^{(LS)} -
\hat\beta_{lj}^{(LS)}|^{-1}$. The use of these weights
(\ref{DefWeights}) was motivated through standardization of design
matrix $Z$ from Section \ref{SectionComp}, in analogy to
standardization of metric predictors. In the following these weights
are also considered, but multiplied by 2. If predictor $x_l$ is
nominal, the factor $(k_l + 1)^{-1}$ is necessary to ensure that
penalty $J_l(\beta_l)$ in (\ref{DefMultPen}) is of order $k_l$, the
number of (free) dummy coefficients. Without these additional
weights $J_l(\beta_l)$ would be of order $(k_l + 1)k_l$, because the
penalty consists of $(k_l + 1)k_l/2$ terms if no ordinal structure
is assumed. By contrast, if the predictor is ordinal, the penalty is
already of order $k_l$. Hence, the factor $2(k_l + 1)^{-1}$ is
omitted in this case.

In general, if weights $w_{ij}^{(l)} \neq 1$ are included, the model
just has to be parameterized by vector $\tilde\theta = W\theta$,
where $W$ is a diagonal matrix with diagonal elements
$w_{ij}^{(l)}$. That means the (centered) design matrix needs to be
multiplied by $W^{-1}$.

\subsection{Refitting procedures}
The most attractive features of the methods described above are
variable selection and clustering. However, due to penalization,
estimates are obviously biased. In the usual ANOVA case, this is not
a problem, since the focus is on the identification of differences,
and not on quantification. In our case---as in regression analysis
in general---we are also interested in parameter estimation and
prediction accuracy. In order to reduce the bias, refitting
procedures have been proposed by several authors, for example, by
\citet{Efronetal2004} under the name ``Lars-OLS hybrid,'' or by
\citet{CanTao2007} as ``Gauss-Dantzig Selector.'' In our setting,
that means that the penalty in (\ref{DefQp}) is only used for
variable selection and clustering. After the identification of
relevant predictors and clusters, parameters are refitted by
ordinary least squares. If variable selection and clustering are
based on the already mentioned adaptive weights, asymptotic behavior
is obtained which is similar to the nonrefitting case; for details,
see the remarks on Proposition \ref{PropCat} in the
\hyperref[app]{Appendix}. However, before we apply the refitting method to
the rent data (where $n < \infty$), its effect is also tested in
simulation studies (see Section \ref{SectionSim2}).

\section{Numerical experiments}

Before applying the presented methodology to the Munich rent
standard data in Section \ref{SectionRent}, the different approaches
are tested and some characteristics are investigated in simulation
studies.

\begin{figure}[b]

\includegraphics{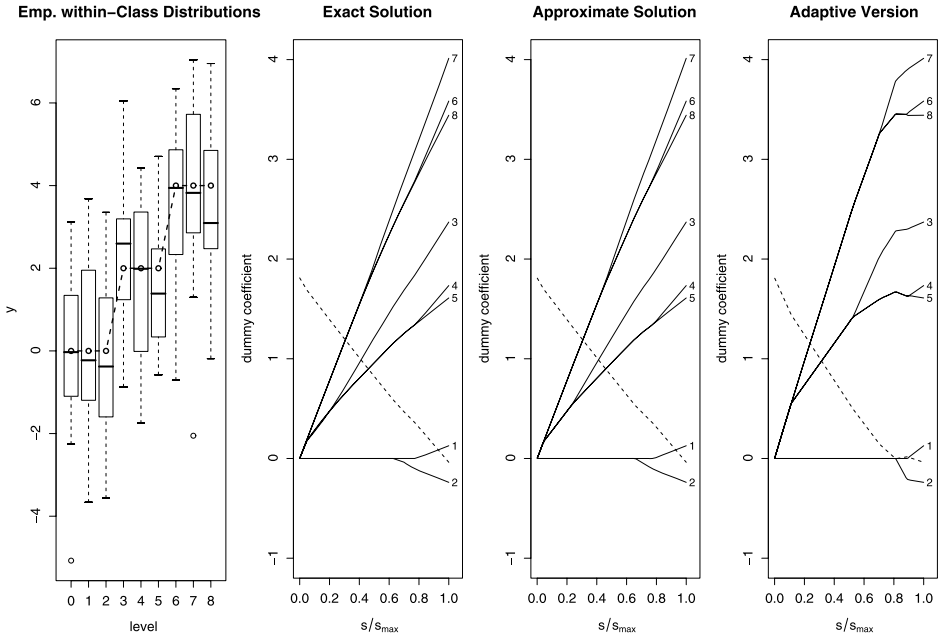}

\caption{Empirical within-class distributions
\textup{(left)}, exact and approximate coefficient paths \textup{(middle)}, as well as
results of the adaptive version \textup{(right)}; constant $\alpha$ is marked
by the dashed line.}
\end{figure}\label{FigureSim1}

\subsection{An illustrative example}
In the first simulation scenario only one predictor and a balanced
design are considered with 20 (independent) observations in each of
$i=0,\ldots,8$ classes. In class $i$ the response is
N($\mu_i$, 4)-distributed, where the means form three distinct groups
of categories, that is, $\mu_0=\mu_1=\mu_2$, $\mu_3=\mu_4=\mu_5$,
$\mu_6=\mu_7=\mu_8$.  Figure \ref{FigureSim1} (left) shows empirical
distributions as well as the true $\mu_i$, which are marked by
dashed lines.
Moreover, exact and approximate paths of dummy coefficients (middle)
are shown, where the nonadaptive version of penalty $J(\beta)$ is
employed. That means the weighting term $|\hat\beta_{i}^{(LS)} -
\hat\beta_{j}^{(LS)}|^{-1}$ is omitted. Since there is only one
predictor and the design is balanced, simple weights $w_{ij} = 1$ can
be used. The \textit{x}-axis indicates $s/s_{\max}$, the ratio of actual and
maximal $s$ value. The latter results in the ordinary least squares
(OLS) estimate. With decreasing $s$ (or increasing penalty
$\lambda$), categories are successively grouped together. First,
classes with the same true mean are grouped as desired; for $s=0$
the model finally consists of the intercept only---the empirical
mean of $y$. For the approximation, $\sqrt{\gamma} = 10^5$ has been
chosen. It is hard to see any difference between approximate and
exact solution. Indeed, for $s/s_{\max} \ge 10^{-3}$, precision
$\Delta_{\gamma,\lambda} < 10^{-17}$ is obtained. Also in the case
of the ``exact'' solution, restrictions are just ``numerically'' met. In
the given example precision of the ``exact'' solution is about
$10^{-18}$ (or better), which is quite close to the ``approximate''
solution. So in the following, only approximate estimates are used.

\begin{figure}[b]

\includegraphics{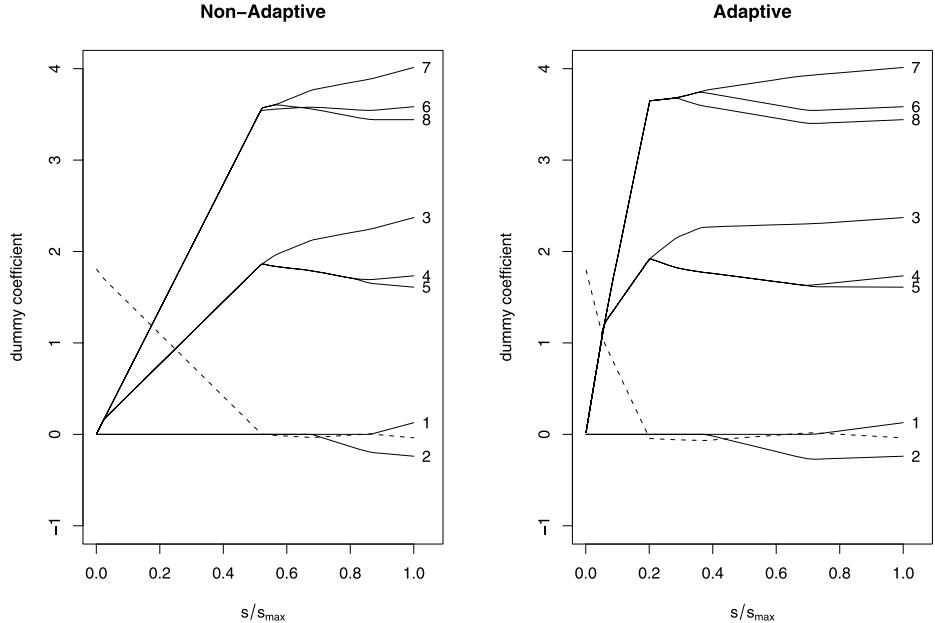}

\caption{Paths of dummy coefficients for data as
in Figure \protect\ref{FigureSim1}, but assuming an ordinal class structure,
nonadaptive \textup{(left)} and adaptive \textup{(right)} version; constant $\alpha$
is marked by the dashed line.}\label{FigureSim2}
\end{figure}

In the right panel of Figure \ref{FigureSim1}, the results of the
adaptive version which uses the additional weights $w_{ij} =
|\hat\beta_{i}^{(LS)} - \hat\beta_{j}^{(LS)}|^{-1}$ are shown.
Grouping is quite good, and compared to the nonadaptive version,
bias toward zero is much smaller at the point of perfect
grouping.

In a second scenario, settings and data visualized in Figure
\ref{FigureSim1} (left) are considered again, but now it is assumed
that class labels have an ordinal structure. Hence, penalty
(\ref{DefPenOrd}) is employed. Resulting paths of dummy coefficients
are plotted in Figure \ref{FigureSim2}. Even for the nonadaptive
version (left), grouping is quite good. Moreover, before optimal
grouping is reached, bias toward zero seems to be quite low. Of
course, assuming an ordinal class structure, which is actually given
because all categories with truly equal coefficients are groups of
neighbors, makes the estimation problem easier.

\subsection{Comparison of methods}\label{SectionSim2}
For the comparison of different methods a setting with 8 predictors
is considered---4 nominal and 4 ordinal factors. For both types of
variables we use two factors with 8 categories and two with 4, of
which in each case only one is relevant. The true nonzero dummy
coefficient vectors are $(0,1,1,1,1,-2,-2)^T$ and $(0,2,2)^T$ for
the nominal predictors, and $(0,1,1,2,2,4,4)^T$ and $(0,-2,-2)^T$
for the ordinal predictors (constant $\alpha = 1$). A training data
set with $n = 500$ (independent) observations is generated according
to the classical linear model with standard normal error $\epsilon$.
The vectors of marginal a priori class probabilities are
$(0.1,0.1,0.2,0.05,0.2,0.1,0.2,0.05)^T$ and $(0.1,0.4,0.2,0.3)^T$
for 8-level and 4-level factors, respectively. The coefficient
vector is estimated by the proposed method, using adaptive as well
as nonadaptive weights. In addition, the effect of taking into
account marginal class frequencies $n_i^{(l)}$ is investigated, which
means we check what happens if $((n_i^{(l)} + n_j^{(l)})/n)^{1/2}$
is omitted in (\ref{DefWeights}). Moreover, refitting is tested (as
already mentioned), that is,~the penalization is only used for variable
selection and clustering. After the identification of relevant
predictors and clusters, parameters are refitted by ordinary least
squares.

For the determination of the right penalty $\lambda$, resp.~$s$
value, we use 5-fold cross-validation. Of course, any information
criterion like AIC or BIC could also be employed. For the latter
some measure of model-complexity is needed. In analogy to the Fused
Lasso [\citet{TibSauRosZhuKni2005}], the degrees of freedom of a
model can be estimated by
\[
\widehat{\mathrm{df}} = 1 + \sum_{l=1}^p k_l^\ast,
\]
where $k_l^\ast$ denotes the number of unique nonzero dummy
coefficients of predictor $x_l$ and the 1 accounts for the intercept.

\begin{figure}

\includegraphics{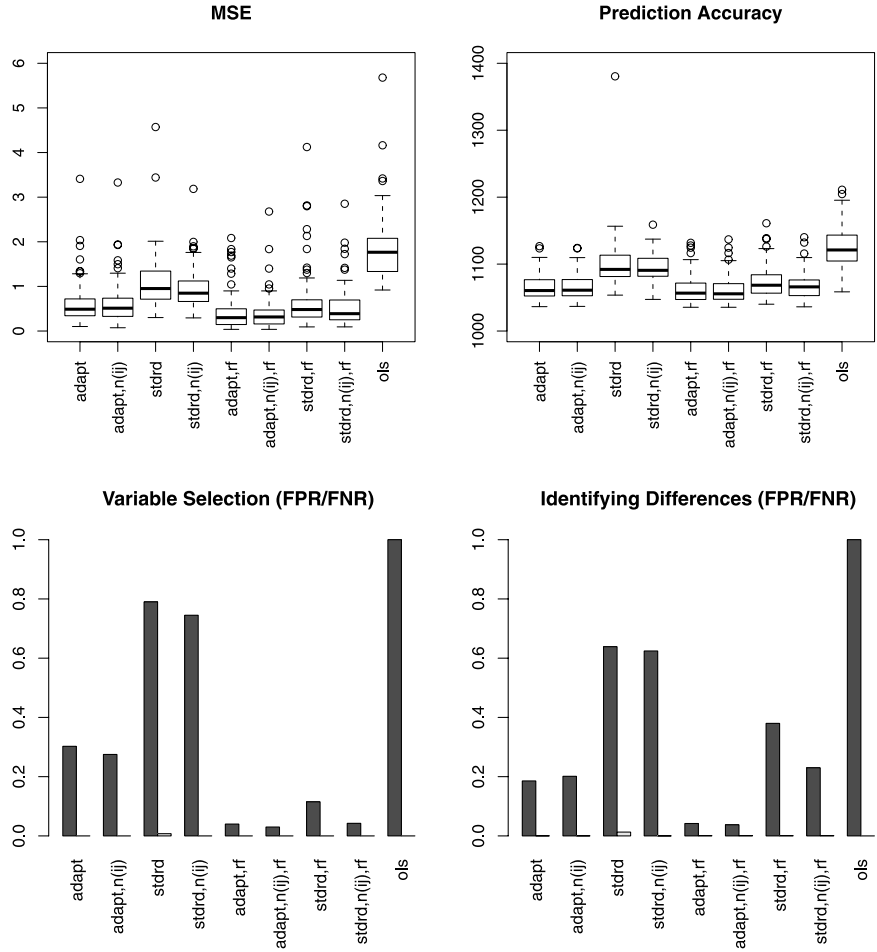}

\caption{Evaluation of adaptive and nonadaptive
(standard) as well as refitting (rf) approaches, taking into account
class sizes ($n_i$, $n_j$) or not, for comparison the results
for the ordinary least squares (ols) estimator are also given; considered
are the mean squared error of parameter estimate, prediction
accuracy and false positive/negative rates (FPR/FNR) concerning
variable selection and identification of relevant differences
(i.e., clustering) of dummy coefficients.}\label{FigureSim3}
\end{figure}

After estimation of coefficient vector $\beta$, the result is
compared to the true parameters. The MSE is computed, as well as
False Positive and False Negative Rates (FPR/FNR) concerning
variable selection and clustering. As far as variable selection is
concerned, ``false positive'' means that any dummy coefficient of a
pure noise factor is set to nonzero; if clustering is considered,
it means that a difference within a nonnoise factor which is truly
zero is set to nonzero. By contrast, ``false negative'' means that
all dummy coefficients of a truly relevant factor are set to zero,
or that a truly nonzero difference is set to zero, respectively.
Figure \ref{FigureSim3} shows the results for 100 simulation runs;
labels are defined in Table \ref{TableSimLabels}.

\begin{table}
\caption{\label{TableSimLabels}Definition of labels used in Figures
\protect\ref{FigureSim3} and \protect\ref{FigureSim4}}
\begin{tabular}{@{}ll@{}}
\hline
adapt & Adaptive version, i.e.,~weighting terms
$|\hat\beta_{i}^{(LS)} - \hat\beta_{j}^{(LS)}|^{-1}$ are used\\
stdrd & Standard (nonadaptive) version, i.e., terms\\
& $|\hat\beta_{i}^{(LS)} - \hat\beta_{j}^{(LS)}|^{-1}$ are omitted\\
n(ij) & Marginal class frequencies are taken into account,\\
& i.e.,~$((n_i^{(l)} + n_j^{(l)})/n)^{1/2}$ are used in (\ref{DefWeights})\\
rf & Refitting was performed\\
\hline
\end{tabular}
\end{table}

In addition to the MSE and FPR/FNR, an independent test set of 1000
observations is generated and prediction accuracies are reported in
terms of the mean squared error of prediction. For comparison the performance of the ordinary least squares (OLS) estimate is also
given. MSE and prediction accuracy are shown as boxplots to give an
idea of variability; FPR (dark gray) and FNR (light-colored) are
averaged over all simulation runs. It is seen that all methods are
superior to the OLS. Concerning FPR and FNR, differences between
pure adaptive/nonadaptive approaches and refitting are caused by the fact that
not necessarily the same models are selected, because in
cross-validation already refitted coefficients are used.

As already illustrated by \citet{BonRei2009} and supported by
Propositions \ref{PropCat} and \ref{PropOrd} in the
\hyperref[app]{Appendix}, selection and grouping characteristics of the
adaptive version are quite good---at least compared with the
standard approach. Also, accuracies of parameter estimates and
prediction of the adaptive version are very high in our simulation
study. Via refitting, they can only be slightly improved. In the case
of standard weights, the improvement is much more distinct. However,
the most important effect of refitting is on variable selection and
clustering---in both the adaptive and the nonadaptive case. It can
be seen that via refitting error rates are enormously diminished---concerning false variable selection as well as clustering. This
finding can be explained by the bias which is caused by shrinking.
If tuning parameters are determined via cross-validation (as done
here), with refitting the chosen penalty parameter $\lambda$ may be
higher than without, because in the latter case a higher penalty
directly results in a higher bias which may deteriorate prediction
accuracy on the test fold. Since in the case of refitting the
penalty is only used for selection purposes, a higher value does not
necessarily cause higher coefficient shrinkage and bias. Apparently,
however, in many of our simulated cases a higher penalty would have
been necessary to obtain accurate variable selection and grouping.

\begin{figure}

\includegraphics{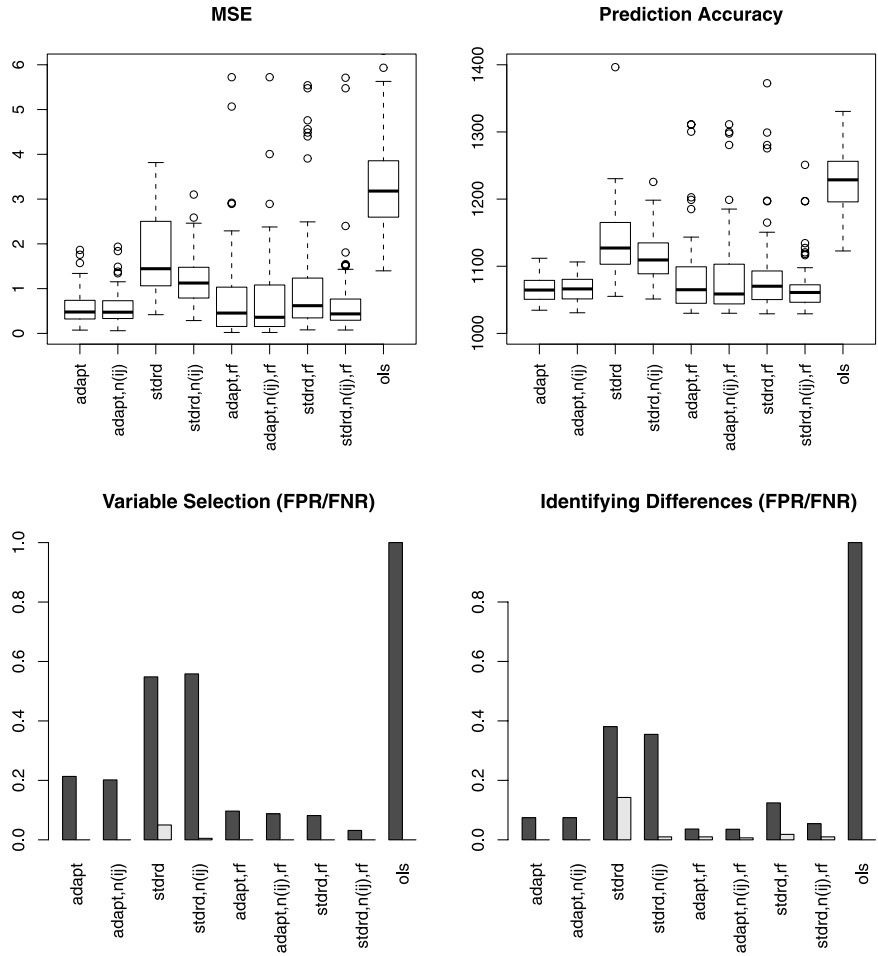}

\caption{Evaluation of different approaches in the
presence of many noise variables: adaptive and nonadaptive
(standard) as well as refitting (rf), taking into account class
sizes ($n_i$, $n_j$) or not, for comparison also the ordinary least
squares (ols) estimator; considered are the mean squared error of
parameter estimate, prediction accuracy and false positive/negative
rates (FPR/FNR) concerning variable selection and identification of
relevant differences (i.e., clustering) of dummy coefficients.}\label{FigureSim4}
\end{figure}

In a modified scenario further noise variables are included, 4
nominal and 4 ordinal, each with 6 levels and constant marginal a
priori class probabilities. Qualitatively, results (shown in Figure
\ref{FigureSim4}) are similar to those obtained before. However,
since the number of independent variables has been considerably
increased, the performance of the ordinary least squares estimates
is even worse than before. This also explains why (in the adaptive
case) the MSE and prediction accuracies cannot be really improved by
OLS refitting, and why in the case of refitting variability is
higher. Nevertheless, variable selection and clustering results are
still distinctly better if refitting is done.

As an overall result, it can be stated that, given a regression
problem, refitting has the potential to distinctly improve selection
and clustering results in the $n < \infty$ case, while providing
accurate parameter estimates (if $n$ is not too small compared to
$p$). Therefore, it can be assumed to be a suitable approach for our
regression analysis. Moreover, taking into account marginal class
frequencies seems to (slightly) improve estimation results.

\section{Regularized analysis of Munich rent standard data}\label{SectionRent}
For the estimation of regression coefficients with predictors from
Table \ref{TableRental1}, we consider the approaches which performed
best in the previous section; more concrete, both the adaptive as
well as the standard (nonadaptive) version remain candidates, but
each with refitting only and taking marginal class frequencies into
account. In the following we first analyze the data and then
evaluate the performance of the approach (using the rent data)
comparing it to ordinary least squares and Group Lasso estimates,
which do not provide variable selection and/or clustering of
categories.

\subsection{Data analysis}
In the considered application more than 2000 observations are
available for the estimation of 58 regression parameters. Thus, OLS
estimation works, and (in the light of the simulation study before)
it is to be expected that refitting distinctly improves estimation
accuracy as well as variable selection and clustering performance of
the proposed penalized approach.

\begin{figure}

\includegraphics{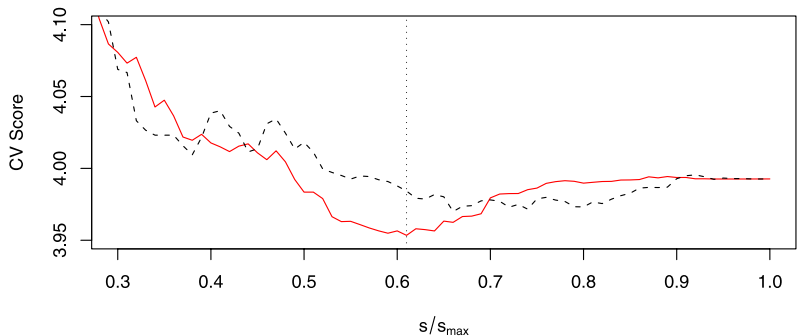}

\caption{Cross-validation score as a function
of $s/s_{\max}$ if refitting with standard (dashed black) or
adaptive (solid red) weights is used for the analysis of Munich rent
standard data.}\label{FigureRental2}
\end{figure}

Figure \ref{FigureRental2} shows the (10-fold) cross-validation
score as a function of $s/s_{\max}$, for the refitted model with
nonadaptive (dashed black) as well as adaptive weights (solid red).
It is seen that penalized estimates, in particular, refitting with
adaptive weights, may improve the ordinary least squares estimate
(i.e.,~$s/s_{\max} = 1$) in terms of prediction accuracy. It is not
surprising that adaptive weights show better performance than
nonadaptive ones, since sample size is high, which means that
ordinary least squares estimates are quite stable, and the latter
decisively influence adaptive weights. So we choose adaptive weights
at cross-validation score minimizing $s/s_{\max} = 0.61$ (marked by
dotted line in Figure \ref{FigureRental2}).
The estimated regression coefficients are given in Table
\ref{TableRental2}.
There is no predictor which is completely excluded from the model.
However, some categories of nominal and ordinal predictors are
clustered, for example, houses constructed in the 1930s and 1940s, or
urban districts 14, 16, 22 and 24. It is interesting that rents of
houses constructed shortly before the Second World War and those
constructed within or shortly after the war do not substantially
differ.

\begin{table}
\caption{\label{TableRental2}Estimated regression coefficients for
Munich rent standard data using adaptive weights with refitting, and
(cross-validation score minimizing) $s/s_{\max} = 0.61$}
\begin{tabular*}{\textwidth}{@{\extracolsep{\fill}}lcc@{}}
  \hline
\textbf{Predictor} & \textbf{Label} & \textbf{Coefficient} \\
  \hline
Intercept & & $\,$12.597 \\[5pt]
Urban district & 14, 16, 22, 24 & $-$1.931 \\
  & 11, 23 & $-$1.719 \\
  & 7 & $-1.622$ \\
  & 8, 10, 15, 17, 19, 20, 21, 25 & $-$1.361 \\
  & 6 & $-1.061$ \\
  & 9 & $-0.960$ \\
  & 13 & $-0.886$ \\
  & 2, 4, 5, 12, 18 & $-$0.671 \\
  & 3 & $-0.403$ \\[5pt]
Year of construction & 1920s & $-$1.244 \\
  & 1930s, 1940s & $-$0.953 \\
  & 1950s & $-0.322$ \\
  & 1960s & \phantom{$-$}0.073 \\
  & 1970s & \phantom{$-$}0.325 \\
  & 1980s & \phantom{$-$}1.121 \\
  & 1990s, 2000s & \phantom{$-$}1.624 \\[5pt]
Number of rooms & 4, 5, 6 & $-$0.502 \\
  & 3 & $-0.180$ \\
  & 2 & \phantom{$-$}0.000 \\[5pt]
Quality of residential area & good & \phantom{$-$}0.373 \\
 & excellent & \phantom{$-$}1.444 \\[5pt]
Floor space $(\mbox{m}^2)$ & $[140,\infty)$ & $-$4.710 \\
 & $[90,100)$, $[100,110)$, $[110,120)$, & \\
 & $[120,130)$, $[130,140)$ & $-$3.688 \\
 & $[60,70)$, $[70,80)$, $[80,90)$ & $-$3.443 \\
 & $[50,60)$ & $-$3.177 \\
 & $[40,50)$ & $-$2.838 \\
 & $[30,40)$ & $-$1.733 \\[5pt]
Hot water supply & no & $-$2.001 \\
Central heating & no & $-$1.319 \\
Tiled bathroom & no & $-$0.562 \\
Suppl.~equipment in bathroom & yes & \phantom{$-$}0.506 \\
Well equipped kitchen & yes & \phantom{$-$}1.207 \\
   \hline
\end{tabular*}
\end{table}

\begin{figure}

\includegraphics{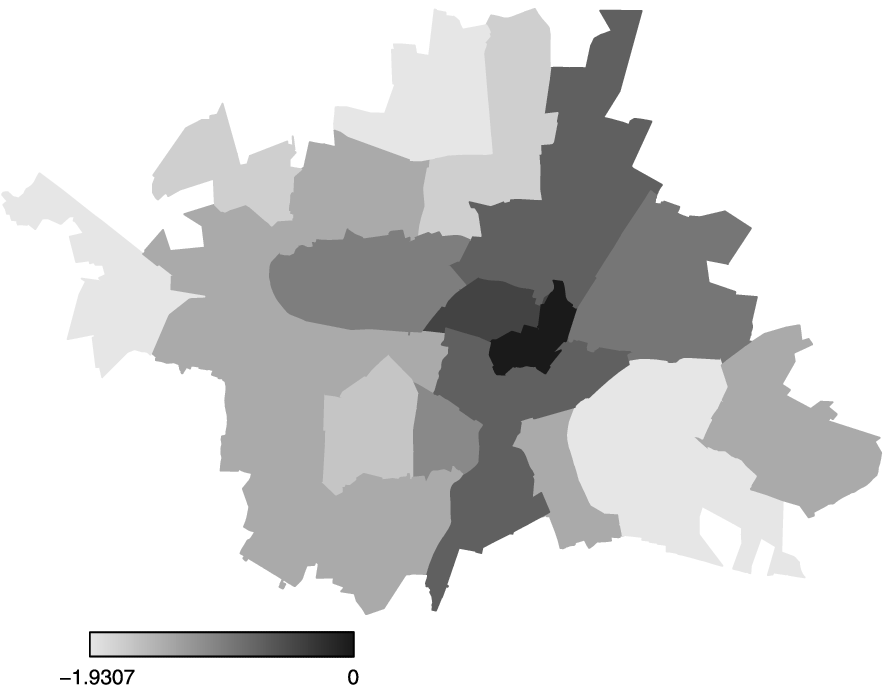}

\caption{Map of Munich indicating clusters of
urban districts; colors correspond to estimated dummy coefficients
from Table \protect\ref{TableRental2}.}\label{FigureRental3a}
\end{figure}

The biggest cluster, which contains 8 categories, is formed within
the 25 districts. A map of Munich with color coded clusters (Figure
\ref{FigureRental3a}) illustrates the 10 found clusters. The map has
been drawn using functions from R add-on package \texttt{BayesX}
[\citet{KneHeiBreSab2009}].
The most expensive district is the city center. After inspection of
OLS estimates (e.g.,~in Figure \ref{FigureRental1b}), it could be
expected that rather cheap districts 14 and 24 are fused. It was not
clear, however, if they are additionally collapsed with any other
districts, and if so, whether fused with $\{16,22\}$ or $\{11,23\}$.
Based on our regularized analysis, it can now be stated with good
reason that rents in districts 14, 16, 22 and 24 are comparatively
low and do not substantially differ, which is in agreement with
judgements from experts and feelings of laymen, because Munich's
deprived areas are primarily located in these (nonadjacent)
districts. The cluster that contains district 12, however, partly
contradicts experiences of experts and tenants. The problem is that
this district is very large and reaches from the city center to the
outskirts in the north. So very expensive flats which are close to
the city center are put together with cheaper ones on the outskirts.
But, on average, rents are rather high in this district, which
causes it to be clustered with other expensive but more
homogeneous areas.
In an ordinary least squares model, district 12 is even identified
as belonging to the three most expensive districts (see also Figures
\ref{FigureRental1a} and \ref{FigureRental1b}). Penalized estimation
ranks it only among the top seven. But it should be noted that in
the final regression model there is also an ordinal predictor
included which indicates the quality of the residential area and
allows for further discrimination between flats which are located in
the same district.

Not surprisingly, rent per square meter goes down if the number of
rooms increases. Between four, five or more rooms, however, no
relevant differences are identified. Flats with two rooms are fused
with the reference category, since the corresponding dummy
coefficient is set to zero. The fact that no differences between
flats with one and two rooms are found is caused by the inclusion of
floor space into the model. Existing differences are obviously
modeled via the variable which directly measures the flat's size,
with the effect that for larger flats the rent per square meter is
lower. Starting with small apartments, the decrease of rents is
quite apparent (between ca.~20 and 60$\mbox{ m}^2$), then it is much
slower. Between 90 and 140$\mbox{ m}^2$, for example, no differences
are identified with respect to rent per square meter. The fact that
the covariate which indicates the number of rooms is not completely
excluded from the model, although the flat's floor space is also
considered, shows that there are dependencies between rent and the
number of rooms which do not only refer to the flat's size. If
covariate floor space is held constant, but the number of rooms is
increased, rents tend to go down.

All in all, the selected model has 32 degrees of freedom, that is,~32
unique nonzero coefficients (including the intercept), which means
that the complexity of the unrestricted model (58 df) is reduced by
about 45\%.

\begin{figure}

\includegraphics{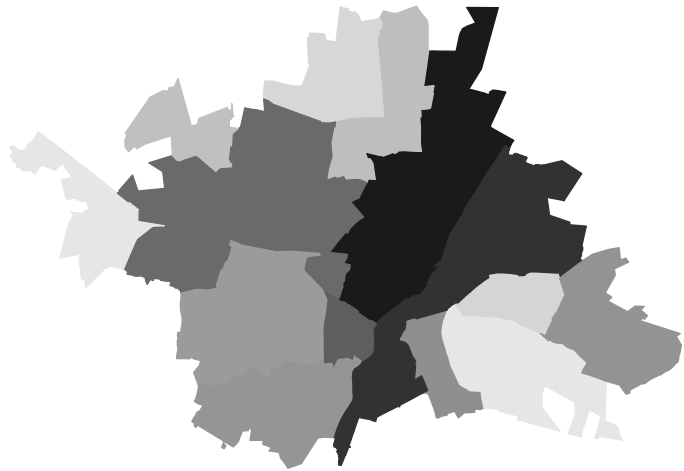}

\caption{Map of Munich indicating clusters of
urban districts, if just differences between dummy coefficients of
neighboring districts are penalized.}\label{FigureRental3b}
\end{figure}

\subsection{Using spatial information}
A possible alternative to treating the urban district as a nominal
predictor is to include geographical information. One can use
distance measures or neighborhood effects when looking for clusters.
A simple approach we used is to penalize---in analogy to the
ordinal predictor case---only  differences between dummy
coefficients of neighboring districts. In Figure
\ref{FigureRental3b} a map of Munich is shown which results from
such neighborhood penalization. One problem with this map is that
district 12 (which reaches from the center to the north) is now
fused with three expensive adjacent districts in the center. We also
fitted a more advanced neighborhood weighting scheme, which uses the
length of the boundary between the corresponding districts as
weights. Then the difference between district 12 and a neighboring
(and cheap) district in the north would get some higher weight.
However, even that modification does not solve the second problem
linked with that kind of spatial information based regularization:
Two nonadjacent districts will not be fused if they are not also
fused with a whole set of districts building a chain that connects
them. In Figure \ref{FigureRental3b} the two light-colored districts
in the west and southeast (22 and 16) seem quite similar. In
contrast to Figure \ref{FigureRental3a} and Table
\ref{TableRental2}, however, they are not fused. The corresponding
difference of dummy coefficients is about 0.007---close to, but not
exactly zero. Generally speaking, districts which are not
neighbors may also be quite similar. Therefore, fusion of such districts
should be possible, too. Hence, we prefer an approach like our
initial modeling where all pairwise differences of districts' dummy
coefficients have been penalized.

A more general procedure to include spatial information is to
incorporate this information into the weights $w_{ij}$ in
(\ref{DefPenCat}). For that purpose weights may be additionally
multiplied by factors $\zeta_{ij}$, where $\zeta_{ij}$ contains
spatial information. As long as $0 < \zeta_{ij} < \infty$ for all
$i,j$, consistency as given in Proposition \ref{PropCat} is not
affected, and all pairwise differences are still penalized as
desired in our application. Factor $\zeta_{ij}$ can, for example, be
defined as a decreasing function of the distance between districts
$i$ and $j$. A special case of such an approach is to penalize only
differences of neighboring districts as already done before. This,
however, does not guaranty $\zeta_{ij} > 0$ for all $i,j$, and did
not produce good results in our application (as shown above).
Furthermore, it seems sensible to assume that differences
(concerning rents) between the city center and the outskirts tend to
be larger than differences between outskirts in the west and the
east of a city. So for defining $\zeta_{ij}$ we may use the
information whether a district is rather central or peripheral. If
$\varsigma_i$ denotes the distance of (the center of) district $i$
to the city center (in km), we define
\[
\zeta_{ij} = K\biggl(\frac{\varsigma_i - \varsigma_j}{h} \biggr),
\]
with a fixed kernel $K$ and bandwidth $h$. For $K$ we use the
Epanechnikov kernel, and $h = 15$ (km), which is roughly the radius
of the smallest circle around the city center which contains the
whole city of Munich. Incorporating spatial information this way,
however, yields exactly the same clustering as already given in
Table \ref{TableRental2}, where urban districts have just been
treated as a nominal predictor. So we can keep interpretations given
above, and will just use the districts' categorial character in the
following. The finding that results do not change if $\zeta_{ij}$
are included is obviously due to the fact that weights are
decisively influenced by the ols terms (see Proposition
\ref{PropCat} in the \hyperref[app]{Appendix}).

\subsection{Evaluation of prediction accuracies and sparsity}
\begin{figure}

\includegraphics{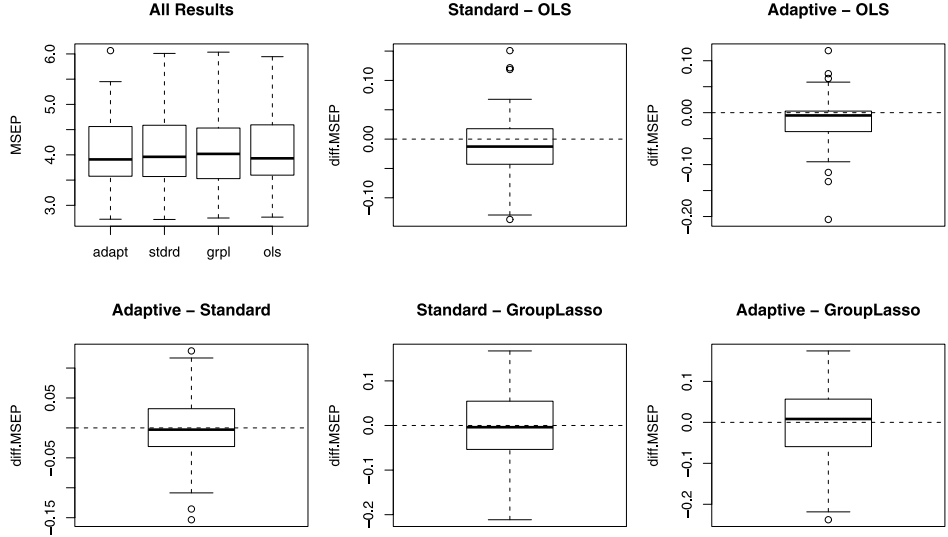}

\caption{Prediction performance of
the refitted adaptive (adapt) as well as nonadaptive (stdrd) sparse
modeling of Munich rent standard data, the Group Lasso (grpl) and
ordinary least squares (ols) fitting; all results \textup{(top left)} as well
as selected pairwise comparisons.}\label{FigureRental4}
\end{figure}

The proposed methods provide clustering of categories, which results
in a sparser model and facilitates interpretation in the considered
application. In order to evaluate their actual prediction accuracies,
we perform repeated random splitting of the data into training and
test sets. That means coefficients are estimated on the training
data (including determination of tuning parameters and weights), and
then used to predict the test data. As test set size we choose 100,
and the procedure is independently repeated 100 times. Results are
shown in Figure \ref{FigureRental4}. Performance is measured in
terms of the mean squared error of prediction (MSEP). We investigate
the refitted adaptive as well as the nonadaptive version of the
presented regularization technique. For comparison, we also give
prediction accuracies for the (most complex) ordinary least squares
model, and for Group Lasso estimates as proposed by
\citet{YuanLin2006} or \citet{MeiGeeBue2008}. In the case of
ordinal predictors, the usual within groups simple ridge penalty is
replaced by a difference penalty as proposed in \citet{GerTut2009e}
and \citet{GerHogObeTut2009}. For practical estimation of Group
Lasso estimates R add-on package \texttt{grplasso}
[\citet{Meier2007}] was used.

The first plot (top left) in Figure \ref{FigureRental4} shows
boxplots of the observed MSEPs for all four methods. It is seen that
all methods perform almost equally. This finding is confirmed by
pairwise comparisons. Since in each iteration MSEPs of different
methods are observed on the same test data, we report pairwise
differences of corresponding MSEPs. Boxplots which tend to be below
zero indicate superior performance of the method which is quoted
first---and vice versa. It just seems that the proposed adaptive
version is slightly superior to the ordinary least squares estimate.
Between the different penalization techniques---the presented
sparse modeling (adaptive/nonadaptive) and the Group Lasso---there
can hardly be observed any difference concerning prediction accuracy
on the rent standard data.

\begin{figure}

\includegraphics{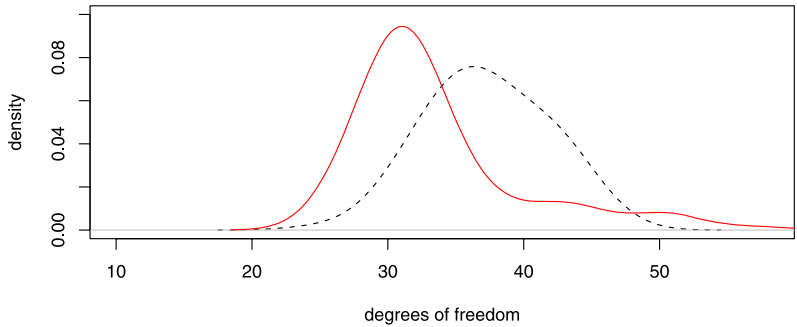}

\caption{Kernel density estimates of chosen
degrees of freedom for the adaptive (solid red) and nonadaptive
(dashed black) model after repeated random splitting of Munich rent
standard data.}\label{FigureRental5}
\end{figure}

It is a quite positive result, however, that prediction accuracies
of the considered methods are almost identical, because sparsity is
the great advantage of the modeling which has been applied above to
analyze the data. While the ordinary least squares model has 58
degrees of freedom, the (refitted adaptive) model which has been
chosen on the basis of all data just has 32 df (see Table
\ref{TableRental2}). In Figure \ref{FigureRental5} we now show
kernel density estimates of the model complexities observed during
random splitting of the data. It is seen that the adaptive models
(solid red) tend to have less degrees of freedom than the
nonadaptive version (dashed black). But also the latter is far away
from the 58 df of the OLS model. Furthermore, the Group Lasso can
only perform variable selection, but no clustering of single
categories. However, in each of the considered random splits all
factors were selected (not shown), which means that none of the
dummy coefficients estimated by the Group Lasso were set to zero.
Hence, with the (via cross-validation) chosen tuning parameters, the
effect of the Group Lasso penalty was just shrinkage/smoothing of
groups of dummy coefficients, but no variable selection. That means
in the case of the analyzed rent standard data the Group Lasso does
not result in a sparser parametrization than the OLS model. In
summary, on the rent data our model can be expected to be as
accurate as competing models, while complexity is distinctly reduced
and interpretability is increased.

\section{Summary and discussion}
We showed how $L_1$-penalization of dummy coefficients can be
employed for sparse modeling of categorial explanatory variables in
multiple linear regression. Depending on the scale level of the
categorial predictor, two types of penalties were investigated. Given
just nominal covariates, all pairwise differences of dummy
coefficients belonging to the same predictor are penalized. If the
variable has ordinal scale level differences of adjacent
coefficients are considered. $L_1$-penalization causes that certain
differences are set to zero. The interpretation is clustering of
categories concerning their influence on the response. In the
analysis of the rent standard data this meant that, for example,~certain
urban districts were identified where rents do not substantially
differ on average. If all dummy coefficients which belong to a
certain predictor are set to zero, the corresponding covariate is
completely removed from the model.

In particular, it was shown that the usually applied (and accurate)
ordinary least squares fitting of rent standard data can be improved
if categorial predictors are adequately penalized. Such improvement
is primarily in terms of interpretability and model complexity. Via
repeated random splitting of the data at hand, it could be shown that
model complexity could be reduced by about 40--50\% while prediction
accuracies did not deteriorate. As simulation studies showed, in
cases of smaller sample sizes estimation and prediction
accuracies can also be distinctly improved via the presented
$L_1$-difference-penalization.

An alternative approach would be to apply clustering methods on ols
estimates, which may give similar results for the considered rent
data (see Figure \ref{FigureRental1b}, though it is not clear, for
example, in which way districts $\{14,24\}$ should be fused with
other districts). However, this would be a two-step procedure and
hence less elegant than a penalty based regularization technique.
Moreover, in case of smaller sample sizes it would severely suffer
from instability of ols estimates.

Though penalization with adaptive weights has some nice asymptotic
properties, simulation studies also showed that in the case of
finite $n$ particularly variable selection and clustering
performance can even be further improved via ordinary least squares
refitting of fused categories. A generalization of refitting is the
so-called relaxed Lasso [\citet{Meinsh2007}], which puts a second
penalty on (dummy) coefficients of fused categories. The
disadvantage of relaxation is the second tuning parameter. In the case
of the Munich rent standard, sample sizes are so high that accurate
(ordinary) least squares estimation is possible, which means that
the second penalty parameter can be omitted.

In the case of ordinal predictors, computation of the proposed
estimator is easily carried out by the \texttt{lars} algorithm
[\citet{Efronetal2004}], since the estimate is just an ordinary
Lasso solution, if independent variables are split-coded. If
predictors are nominal, we showed how procedures designed for
ordinary Lasso problems can also be used to compute approximate
coefficient paths.


\begin{appendix}\label{app}
\section*{Appendix}
\subsection*{Asymptotic properties for the unordered case}
 Let $\theta =
(\theta_{10},\theta_{20},\ldots,\break\theta_{k,k-1})^T$ denote the vector
of pairwise differences $\theta_{ij} = \beta_i - \beta_j$.
Furthermore, let $\mathcal{C} = \{ (i,j) \dvtx \beta_i^\ast \neq
\beta_j^\ast, i > j\}$ denote the set of indices $i > j$
corresponding to differences of (true) dummy coefficients
$\beta_i^\ast$ which are truly nonzero, and $\mathcal{C}_n$ denote
the set corresponding to those difference which are estimated to be
nonzero with sample size $n$, and based on estimate $\hat\beta$
from (\ref{Defbetahat}) with penalty (\ref{DefPenCat}). If
$\theta_\mathcal{C}^\ast$ denotes the true vector of pairwise
differences included in $\mathcal{C}$, $\hat\theta_\mathcal{C}$
denotes the corresponding estimate based on $\hat\beta$, and
$\hat\beta_i^{(LS)}$ the ordinary least squares estimate of
$\beta_i$, then a slightly modified version of Theorem 1 in
\citet{BonRei2009} holds:

\begin{proposition}\label{PropCat}
Suppose $\lambda = \lambda_n$ with $\lambda_n/\sqrt{n} \rightarrow
0$ and $\lambda_n \rightarrow \infty$, and all class-wise sample
sizes $n_i$ satisfy $n_i/n \rightarrow c_i$, where $0 < c_i < 1$.
Then weights $w_{ij} = \phi_{ij}(n)|\hat\beta_i^{(LS)} -
\hat\beta_j^{(LS)}|^{-1}$, with $\phi_{ij}(n) \rightarrow q_{ij}$
$(0 < q_{ij} < \infty)$ $\forall i,j$, ensure that:
\begin{enumerate}
\item[(a)] $\sqrt{n}(\hat\theta_\mathcal{C} - \theta_\mathcal{C}^\ast)
\rightarrow_d N(0,\Sigma)$,
\item[(b)] $\lim_{n \rightarrow \infty} P(\mathcal{C}_n = \mathcal{C}) =
1$.
\end{enumerate}
\end{proposition}

 \begin{Rem*}
 The proof closely follows
\citet{Zou2006} and \citet{BonRei2009}, and is given below. The
main differences to \citet{BonRei2009} are that a concrete form of
the dependence on sample size, specified in $\phi_{ij}(n)$, is not
yet fixed, and that $\lambda_n$ is determined by $\lambda_n/\sqrt{n}
\rightarrow 0$ and $\lambda_n \rightarrow \infty$. The latter is
needed for the proof of asymptotic normality, as given in
\citet{Zou2006}. \citet{BonRei2009} used $\lambda_n =
O_p(\sqrt{n})$, which also allows $\lambda_n = 0$ and therefore
cannot yield $\lim_{n \rightarrow \infty} P(\mathcal{C}_n =
\mathcal{C}) = 1$. $\phi_{ij}(n)$ only needs to converge toward a
positive finite value (denoted by $q_{ij}$). Note that the
covariance matrix $\Sigma$ of the asymptotic normal distribution is
singular due to linear dependencies of pairwise differences;
cf.~\citet{BonRei2009}. The concrete form of $\Sigma$ results from
the asymptotic marginal distribution of a set of nonredundant truly
nonzero differences as specified in the proof.

Due to the (additive) form of the penalty (\ref{DefMultPen}),
theoretic results from above directly generalize to the case of
multiple categorial inputs, given the number $p$ of predictors and
the number $k_l$ of levels of each predictor $x_l$ are fixed.

Simple consistency
$\lim_{n\rightarrow\infty}P(\Vert\hat\beta-\beta^\ast\Vert^2 > \varepsilon) =
0$ for all $\varepsilon > 0$ is also reached if $\lambda$ is fixed and
$w_{ij} = \phi_{ij}(n)$, with $\phi_{ij}(n) \rightarrow q_{ij}$ $(0
< q_{ij} < \infty)$ $\forall i,j$, is chosen. This behavior is
formally described in Proposition \ref{PropConsis}.

If adaptive weights are used and refitting is applied after the
identification of clusters and relevant variables, asymptotic
behavior is obtained which is comparable to Proposition
\ref{PropCat}. Since clustering and variable selection are directly
based on the penalty with adaptive weights, part (b) of Proposition
\ref{PropCat} is still valid. Asymptotic normality results from
asymptotic normality of the ordinary least squares refit.
\end{Rem*}

\begin{pf*}{Proof of Proposition \ref{PropCat}}\label{ProofCat}
 We first show asymptotic normality,
which closely follows \citet{Zou2006} and \citet{BonRei2009}.
Coefficient vector $\beta$ is represented by $u = \sqrt{n}(\beta -
\beta^\ast)$, that is,~$\beta = \beta^\ast + u/\sqrt{n}$, where
$\beta^\ast$ denotes the true coefficient vector. Then we also have
$\hat\beta = \beta^\ast + \hat{u}/\sqrt{n}$, with
\[
\hat{u} = \mathop{\operatorname{argmin}}_{u}\Psi_n(u),
\]
 where
\[
\Psi_n(u) = \biggl(y - X\biggl(\beta^\ast+\frac{u}{\sqrt{n}}\biggr)\biggr)^T
\biggl(y - X\biggl(\beta^\ast+\frac{u}{\sqrt{n}}\biggr)\biggr) +
\frac{\lambda_n}{\sqrt{n}}J(u),
\]
with
\begin{eqnarray*}
J(u) &=& \sum_{i>j; i,j\neq 0}
\sqrt{n}\frac{\phi_{ij}(n)}{|\hat\beta_i^{(LS)}-\hat\beta_j^{(LS)}|}\biggl|\beta_i^\ast-\beta_j^\ast
+ \frac{u_i - u_j}{\sqrt{n}}\biggr|\\
&&{}+ \sum_{i > 0}
\sqrt{n}\frac{\phi_{i0}(n)}{|\hat\beta_i^{(LS)}|}\biggl|\beta_i^\ast
+ \frac{u_i}{\sqrt{n}} \biggr|.
\end{eqnarray*}
Furthermore, since $y - X\beta^\ast = \epsilon$, we have $\Psi_n(u)
- \Psi_n(0) = V_n(u)$, where
\[
V_n(u) = u^T\biggl(\frac{1}{n}X^TX\biggr)u - 2\frac{\epsilon^TX}{\sqrt{n}}u + \frac{\lambda_n}{\sqrt{n}}\widetilde{J}(u),
\]
with
\begin{eqnarray*}
\widetilde{J}(u) &=& \sum_{i>j; i,j\neq 0}
\sqrt{n}\frac{\phi_{ij}(n)}{|\hat\beta_i^{(LS)}-\hat\beta_j^{(LS)}|}\biggl(\biggl|\beta_i^\ast-\beta_j^\ast
+ \frac{u_i - u_j}{\sqrt{n}}\biggr| - |\beta_i^\ast-\beta_j^\ast| \biggr)\\
&&{}+ \sum_{i > 0}
\sqrt{n}\frac{\phi_{i0}(n)}{|\hat\beta_i^{(LS)}|}\biggl(\biggl|\beta_i^\ast
+ \frac{u_i}{\sqrt{n}}\biggr| - |\beta_i^\ast|\biggr).
\end{eqnarray*}
As given in \citet{Zou2006}, we will consider the limit behavior of
$(\lambda_n/\sqrt{n})\widetilde{J}(u)$. If $\beta_i^\ast \neq 0$,
then
\[
\bigl|\hat\beta_i^{(LS)}\bigr| \rightarrow_{\!p}\, |\beta_i^\ast|,\quad \mbox{and}\quad
\sqrt{n}\biggl(\biggl|\beta_i^\ast + \frac{u_i}{\sqrt{n}}\biggr| -
|\beta_i^\ast| \biggr) = u_i \mbox{sgn}(\beta_i^\ast);
\]
and similarly, if $\beta_i^\ast \neq \beta_j^\ast$,
\begin{eqnarray*}
&&\hspace*{106pt}\bigl|\hat\beta_i^{(LS)}-\hat\beta_j^{(LS)}\bigr| \rightarrow_{\!p}\, |\beta_i^\ast
- \beta_j^\ast|,\quad \mbox{and}\\
&&\sqrt{n}\biggl(\biggl|\beta_i^\ast-\beta_j^\ast
+ \frac{u_i - u_j}{\sqrt{n}}\biggr| - |\beta_i^\ast-\beta_j^\ast|
\biggr) = (u_i - u_j) \mbox{sgn}(\beta_i^\ast-\beta_j^\ast).
\end{eqnarray*}
Since by assumption $\phi_{ij}(n) \rightarrow q_{ij}$ $(0 < q_{ij} <
\infty)$ and $\lambda_n/\sqrt{n} \rightarrow 0$, by Slutsky's
theorem, we have
\[
\frac{\lambda_n}{\sqrt{n}}\frac{\phi_{i0}(n)}{|\hat\beta_i^{(LS)}|}\sqrt{n}\biggl(\biggl|\beta_i^\ast
+ \frac{u_i}{\sqrt{n}}\biggr| - |\beta_i^\ast|\biggr) \rightarrow_{\!p}\,
0,
\]
and
\begin{eqnarray*}
&&\frac{\lambda_n}{\sqrt{n}}\frac{\phi_{ij}(n)}{|\hat\beta_i^{(LS)}-\hat\beta_j^{(LS)}|}
\sqrt{n}\biggl(\biggl|\beta_i^\ast-\beta_j^\ast + \frac{u_i -
u_j}{\sqrt{n}}\biggr| - |\beta_i^\ast-\beta_j^\ast| \biggr)\\
&&\qquad\rightarrow_{\!p}\, 0, \qquad\mbox{ respectively}.
\end{eqnarray*}
This also makes clear that
assumption $\lambda_n = O_p(\sqrt{n})$ is not enough. If
$\beta_i^\ast = 0$ or $\beta_i^\ast = \beta_j^\ast$, however,
\begin{eqnarray*}
\sqrt{n}\biggl(\biggl|\beta_i^\ast + \frac{u_i}{\sqrt{n}}\biggr| -
|\beta_i^\ast| \biggr)& =& |u_i|,\quad \mbox{and}\\
\sqrt{n}\biggl(\biggl|\beta_i^\ast-\beta_j^\ast
+ \frac{u_i - u_j}{\sqrt{n}}\biggr| - |\beta_i^\ast-\beta_j^\ast|
\biggr) &=& |u_i - u_j|,\qquad\mbox{respectively}.
\end{eqnarray*}
Moreover, if
$\beta_i^\ast = 0$ or $\beta_i^\ast = \beta_j^\ast$, due to
$\sqrt{n}$-consistency of the ordinary least squares estimate (which
is ensured by condition $n_i/n \rightarrow c_i$, $0 < c_i < 1\
\forall i$),
\begin{eqnarray*}
\lim_{n\rightarrow\infty}
P\bigl(\sqrt{n}\bigl|\hat\beta_i^{(LS)}\bigr| \le \lambda_n^{1/2}\bigr) &=& 1,\qquad \mbox{respectively,}\\ \lim_{n\rightarrow\infty}
P\bigl(\sqrt{n}\bigl|\hat\beta_i^{(LS)}-\hat\beta_j^{(LS)}\bigr| \le
\lambda_n^{1/2}\bigr) &=& 1,
\end{eqnarray*}
 since $\lambda_n \rightarrow \infty$ by
assumption. Hence,
\begin{eqnarray*}
&&\hspace*{105pt}\frac{\lambda_n}{\sqrt{n}}\frac{\phi_{i0}(n)}{|\hat\beta_i^{(LS)}|}\sqrt{n}\biggl(\biggl|\beta_i^\ast
+ \frac{u_i}{\sqrt{n}}\biggr| - |\beta_i^\ast|\biggr) \rightarrow_{\!p}\,
\infty,\quad \mbox{or}\\
&&\frac{\lambda_n}{\sqrt{n}}\frac{\phi_{ij}(n)}{|\hat\beta_i^{(LS)}-\hat\beta_j^{(LS)}|}
\sqrt{n}\biggl(\biggl|\beta_i^\ast-\beta_j^\ast + \frac{u_i -
u_j}{\sqrt{n}}\biggr| - |\beta_i^\ast-\beta_j^\ast| \biggr)
\rightarrow_{\!p}\, \infty,
\end{eqnarray*}
 if $u_i \neq 0$, resp.~$u_i \neq u_j$. That
means if for any $i,j > 0$ with $\beta_i^\ast = \beta_j^\ast$ or
$\beta_i^\ast = 0$, $u_i \neq u_j$ or $u_i \neq 0$, respectively,
then $(\lambda_n/\sqrt{n})\widetilde{J}(u) \rightarrow_{\!p}\, \infty$.
The rest of the proof of part (a) is almost identical to
\citet{BonRei2009}. Let $X^\ast$ denote the design matrix
corresponding to the correct structure, that is,~columns of dummy
variables with equal coefficients are added and collapsed, and
columns corresponding to zero coefficients are removed. 
Since $\forall i$ $n_i/n \rightarrow c_i$ ($0 < c_i < 1$),
\[
\frac{1}{n}X^{\ast T} X^\ast \rightarrow C > 0 \quad\mbox{and}\quad
\frac{\epsilon^TX^\ast}{\sqrt{n}} \rightarrow_d w, \qquad\mbox{with } w
\sim N(0,\sigma^2C).
\]
 Let $\theta_{\mathcal{C}^c}$ denote the
vector of differences $\theta_{ij} = \beta_i - \beta_j$ which are
truly zero, that is,~not from $\mathcal{C}$, and $u_{\mathcal{C}^c}$ the
subset of entries of $\theta_{\mathcal{C}^c}$ which are part of $u$.
By contrast, $u_{\mathcal{C}}$ denotes the subset of
$\theta_{\mathcal{C}}$ which are in $u$. As given in
\citet{Zou2006}, by Slutsky's theorem, $V_n(u) \rightarrow_d V(u)$
for every $u$, where
\[
V(u) = \cases{
u_{\mathcal{C}}^TCu_{\mathcal{C}} - 2u_{\mathcal{C}}^Tw, &\quad \mbox{if }$\theta_{\mathcal{C}^c} =
0$,\cr
\infty, &\quad \mbox{otherwise.}}
\]
Since $V_n(u)$ is convex and the unique minimum of $V(u)$ is
$(C^{-1}w,0)^T$, we have [cf.~\citet{Zou2006};
\citet{BonRei2009}]
\[
\hat{u}_{\mathcal{C}} \rightarrow_d C^{-1}w\quad\mbox{and}\quad
\hat{u}_{\mathcal{C}^c} \rightarrow_d 0.
\]
Hence,
$\hat{u}_{\mathcal{C}} \rightarrow_d N(0,\sigma^2C^{-1})$. By
changing the reference category, that is,~changing the subset of entries
of $\theta$ which are part of $u$, asymptotic normality can be
proven for all pairwise differences in $\hat\theta_\mathcal{C}$.

To show the consistency part, we first note that $\lim_{n
\rightarrow \infty}P((i,j) \in \mathcal{C}_n) = 1$, if $(i,j) \in
\mathcal{C}$, follows from part (a). We will now show that if $(i,j)
\notin \mathcal{C}$, $\lim_{n \rightarrow \infty}P((i,j) \in
\mathcal{C}_n) = 0$. The proof is a modified version of the one
given by \citet{BonRei2009}. Let $\mathcal{B}_n$ denote the
(nonempty) set of pairs of indices $i>j$ which are in
$\mathcal{C}_n$ but not in $\mathcal{C}$. Then we may choose
reference category $0$ such that $\hat\beta_q = \hat\beta_q -
\hat\beta_0 > 0$ is the largest difference corresponding to indices
from $\mathcal{B}_n$. Moreover, we may order categories such that
$\hat\beta_1\le\cdots\le\hat\beta_z\le0\le\hat\beta_{z+1}\le\cdots\le\hat\beta_k$.
That means estimate $\hat\beta$ from (\ref{Defbetahat}) with penalty
(\ref{DefPenCat}) is equivalent to
\[
\hat\beta =
\mathop{\operatorname{argmin}}_{\{\beta_1\le\cdots\le\beta_z\le0\le\beta_{z+1}\le\cdots\le\beta_k\}}
\{(y-X\beta)^T(y-X\beta) + \lambda_n J(\beta)\},
\]
with
\begin{eqnarray*}
J(\beta) &=& \sum_{i>j; i,j\neq 0} \phi_{ij}(n)
\frac{\beta_i-\beta_j}{|\hat\beta_i^{(LS)}-\hat\beta_j^{(LS)}|}\\
&&{} +
\sum_{i \ge z+1} \phi_{i0}(n)\frac{\beta_i}{|\hat\beta_i^{(LS)}|} -
\sum_{i \le z} \phi_{i0}(n)\frac{\beta_i}{|\hat\beta_i^{(LS)}|}.
\end{eqnarray*}
Since $\hat\beta_q \neq 0$ is assumed, at the solution $\hat\beta$
this optimization criterion is differentiable with respect to
$\beta_q$. We may consider this derivative in a neighborhood of the
solution where coefficients which are set equal remain equal. That
means terms corresponding to pairs of indices which are not in
$\mathcal{C}_n$ can be omitted, since they will vanish in $J(\hat\beta)$. If
$x_q$ denotes the $q$th column of design matrix $X$, due to
differentiability, estimate $\hat\beta$ must satisfy
\[
\frac{Q'_q(\hat\beta)}{\sqrt{n}} =
\frac{2x_q^T(y-X\hat\beta)}{\sqrt{n}} = A_n + D_n,
\]
 with
\[
A_n = \frac{\lambda_n}{\sqrt{n}}\biggl( {\sum_{j<q; (q,j)\in \mathcal{C}}
\frac{\phi_{qj}(n)}{|\hat\beta_q^{(LS)}-\hat\beta_j^{(LS)}|}}  -
{\sum_{i>q; (i,q)\in \mathcal{C}}
\frac{\phi_{iq}(n)}{|\hat\beta_i^{(LS)}-\hat\beta_q^{(LS)}|}}\biggr)
\]
and
\[
D_n = \frac{\lambda_n}{\sqrt{n}} \sum_{j<q; (q,j)\in \mathcal{B}_n}
\frac{\phi_{qj}(n)}{|\hat\beta_q^{(LS)}-\hat\beta_j^{(LS)}|}.
\]
 If
$\beta^\ast$ denotes the true coefficient vector,
$Q'_q(\hat\beta)/\sqrt{n}$ can be written as
\[
\frac{Q'_q(\hat\beta)}{\sqrt{n}} =
\frac{2x_q^T(y-X\hat\beta)}{\sqrt{n}} =
\frac{2x_q^TX\sqrt{n}(\beta^\ast-\hat\beta)}{n} +
\frac{2x_q^T\epsilon}{\sqrt{n}}.
\]
 From part (a) and applying
Slutsky's theorem, we know that $2x_q^TX\sqrt{n}(\beta -
\hat\beta)/n$ has some asymptotic normal distribution with mean
zero, and $2x_q^T\epsilon/\sqrt{n}$ as well (by assumption, and
applying the central limit theorem); cf. \citet{Zou2006}. Hence, for
any $\varepsilon > 0$, we have
\[
\lim_{n \rightarrow \infty}P\bigl(Q'_q(\hat\beta)/\sqrt{n} \le \lambda_n^{1/4} - \varepsilon\bigr) = 1.
\]
Since $\lambda_n/\sqrt{n} \rightarrow 0$, we also know $\exists
\varepsilon > 0$ such that $\lim_{n \rightarrow \infty} P(|A_n| <
\varepsilon) = 1$. By assumption, $\lambda_n \rightarrow \infty$; due
to $\sqrt{n}$-consistency of the ordinary least squares estimate, we
know that
\[
\lim_{n\rightarrow\infty} P\bigl(\sqrt{n}\bigl|\hat\beta_q^{(LS)}-\hat\beta_j^{(LS)}\bigr| \le \lambda_n^{1/2}\bigr) = 1,
\]
if $(q,j)\in \mathcal{B}_n$. Hence,
\[
\lim_{n\rightarrow\infty} P(D_n > \lambda_n^{1/4}) = 1.
\]
As a consequence,
\[
\lim_{n\rightarrow\infty} P\bigl(Q'_q(\hat\beta)/\sqrt{n} = A_n + D_n\bigr) = 0.
\]
That means if $(i,j) \notin \mathcal{C}$, also
\[
\lim_{n\rightarrow\infty} P\bigl((i,j) \in \mathcal{C}_n\bigr) = 0.
\]
\upqed\end{pf*}

\begin{proposition}\label{PropConsis}
Suppose $0 \le \lambda < \infty$ has been fixed, and all class-wise
sample sizes $n_i$ satisfy $n_i/n \rightarrow c_i$, where $0 < c_i <
1$. Then weights $w_{ij} = \phi_{ij}(n)$, with $\phi_{ij}(n)
\rightarrow q_{ij}$ $(0 < q_{ij} < \infty)$ $\forall i,j$, ensure
that estimate $\hat\beta$ from (\ref{Defbetahat}) with penalty
(\ref{DefPenCat}) is consistent,
that is,~$\lim_{n\rightarrow\infty}P(\Vert\hat\beta-\beta^\ast\Vert^2 >
\varepsilon) = 0$ for all $\varepsilon > 0$.
\end{proposition}

 \begin{pf}\label{ProofConsis}
  If $\hat\beta$ minimizes
$Q_p(\beta)$ from (\ref{DefQp}), then it also minimizes
$Q_p(\beta)/n$. The ordinary least squares estimator
$\hat\beta^{(LS)}$ minimizes $Q(\beta) = (y-X\beta)^T(y-X\beta)$,
resp.~$Q(\beta)/n$. Since $Q_p(\hat\beta)/n \rightarrow_{\!p}\,
Q(\hat\beta^{(LS)})/n$ and $Q_p(\hat\beta)/n \rightarrow_{\!p}\,
Q(\hat\beta)/n$, we have $Q(\hat\beta)/n \rightarrow_{\!p}\,
Q(\hat\beta^{(LS)})/n$. Since $\hat\beta^{(LS)}$ is the unique
minimizer of $Q(\beta)/n$, and $Q(\beta)/n$ is convex, we have
$\hat\beta \rightarrow_{\!p}\, \hat\beta^{(LS)}$, and consistency follows
from consistency of the ordinary least squares estimator
$\hat\beta^{(LS)}$, which is ensured by condition $n_i/n \rightarrow
c_i$, with $0 < c_i < 1 \  \forall i$.
\end{pf}

\subsection*{Asymptotic properties for the ordered case} Let
now $\mathcal{C} = \{i > 0\dvtx \beta_i^\ast \neq \beta_{i-1}^\ast\}$
denote the set of indices corresponding to differences of
neighboring (true) dummy coefficients $\beta_i^\ast$ which are truly
nonzero, and again, $\mathcal{C}_n$ denote the set corresponding to
those difference which are estimated to be nonzero, based on
estimate $\hat\beta$ from (\ref{Defbetahat}) with penalty
(\ref{DefPenOrd}). The vector of first differences $\delta_i =
\beta_i - \beta_{i-1}$, $i=1,\ldots,k$, is now denoted as $\delta =
(\delta_1,\ldots,\delta_k)^T$. In analogy to the unordered case,
$\delta_\mathcal{C}^\ast$ denotes the true vector of (first)
differences included in $\mathcal{C}$, and $\hat\delta_\mathcal{C}$
the corresponding estimate. With $\hat\beta_i^{(LS)}$ denoting the
ordinary least squares estimate of $\beta_i$, the following
proposition holds.

\begin{proposition}\label{PropOrd}
Suppose $\lambda = \lambda_n$ with $\lambda_n/\sqrt{n} \rightarrow
0$ and $\lambda_n \rightarrow \infty$, and all class-wise sample
sizes $n_i$ satisfy $n_i/n \rightarrow c_i$, where $0 < c_i < 1$.
Then weights $w_i = \phi_i(n)|\hat\beta_i^{(LS)} -
\hat\beta_{i-1}^{(LS)}|^{-1}$, with $\phi_i(n) \rightarrow q_i$ $(0
< q_i < \infty)$ $\forall i$, ensure that:
\begin{enumerate}
\item[(a)] $\sqrt{n}(\hat\delta_\mathcal{C} - \delta_\mathcal{C}^\ast)
\rightarrow_d N(0,\Sigma)$,
\item[(b)] $\lim_{n \rightarrow \infty} P(\mathcal{C}_n = \mathcal{C}) =
1$.
\end{enumerate}
\end{proposition}

 \begin{Rem*} The proof is a direct application of
Theorem 2 in \citet{Zou2006}, as sketched below. As before in the
unordered case, if $\lambda$ is fixed and $w_{i} = \phi_{i}(n)$,
with $\phi_{i}(n) \rightarrow q_{i}$ $(0 < q_{i} < \infty)$ $\forall
i,j$, simple consistency $\lim_{n\rightarrow\infty}
P(\Vert\hat\beta-\beta^\ast\Vert^2 > \varepsilon) = 0$ for all $\varepsilon > 0$
is reached. The proof is completely analogue to the proof of
Proposition \ref{PropConsis} before.
\end{Rem*}

 \begin{pf*}{Proof of Proposition \ref{PropOrd}}\label{ProofOrd}
  In Section \ref{SectionComp} it has been shown that the proposed estimate given an ordinal class
structure is equivalent to a Lasso type estimate, if ordinal
predictors are split-coded. That means since $\phi_i(n) \rightarrow
q_i$ $(0 < q_i < \infty)$ $\forall i$ by assumption, and employing
Slutsky's theorem (the proof of), Theorem~2 about the adaptive Lasso
by \citet{Zou2006} can be directly applied. Condition $n_i/n
\rightarrow c_i$, with $0 < c_i < 1\  \forall i$, ensures that the
ordinary least squares estimate is $\sqrt{n}$-consistent.
\end{pf*}

\subsection*{Precision of the approximate solution}
\begin{proposition}\label{PropPrecision}
If restriction $\theta_{ij} = \theta_{i0} - \theta_{j0}$ is
represented by $A\theta = 0$, define $\hat\theta_{\gamma,\lambda} =
\mathop{\operatorname{argmin}}_{\theta} \{(y-Z\theta)^T(y-Z\theta) +
\gamma(A\hat\theta)^TA\hat\theta + \lambda|\theta|\}$, where $\theta
= (\theta_{10},\ldots,\theta_{k,k-1})^T$ and $|\theta| = \sum_{i>j}
|\theta_{ij}|$. Then with $\gamma > 0$ and $\lambda \ge 0$,
$\Delta_{\gamma,\lambda} =
(A\hat\theta_{\gamma,\lambda})^TA\hat\theta_{\gamma,\lambda}$ is
bounded above by
\[\Delta_{\gamma,\lambda} \le \frac{\lambda
(|\hat\theta^{(LS)}| - |\hat\theta_{0,\lambda}|)}{\gamma},\] where
$\hat\theta^{(LS)}$ denotes the least squares estimate (i.e.,
$\lambda = 0$) where $A\hat\theta^{(LS)}=0$ holds.
\end{proposition}

 \begin{pf}\label{ProofPrecision}
  Obviously, for all
$\gamma > 0$ and $\lambda \ge 0$,
\begin{eqnarray*}
&&(y-Z\hat\theta_{\gamma,\lambda})^T(y-Z\hat\theta_{\gamma,\lambda})
+ \lambda|\hat\theta_{\gamma,\lambda}| +
\gamma\Delta_{\gamma,\lambda}\\
&&\qquad \le
\bigl(y-Z\hat\theta^{(LS)}\bigr)^T\bigl(y-Z\hat\theta^{(LS)}\bigr) +
\lambda\bigl|\hat\theta^{(LS)}\bigr|.
\end{eqnarray*}
 Since also
\[
(y-Z\hat\theta_{0,\lambda})^T(y-Z\hat\theta_{0,\lambda})
+ \lambda|\hat\theta_{0,\lambda}| \le
(y-Z\hat\theta_{\gamma,\lambda})^T(y-Z\hat\theta_{\gamma,\lambda}) +
\lambda|\hat\theta_{\gamma,\lambda}|,
\]
 and
\[
(y-Z\hat\theta_{0,\lambda})^T(y-Z\hat\theta_{0,\lambda}) \ge
\bigl(y-Z\hat\theta^{(LS)}\bigr)^T\bigl(y-Z\hat\theta^{(LS)}\bigr),
\]
 we have
\[
\gamma\Delta_{\gamma,\lambda} \le \lambda\bigl(\bigl|\hat\theta^{(LS)}\bigr| -|\hat\theta_{0,\lambda}|\bigr).
\]
\upqed\end{pf}
\end{appendix}

\printaddresses

\end{document}